\documentclass[bimj,fleqn]{w-art}
\usepackage{times}
\usepackage{w-thm}

\usepackage{makecell}
\usepackage{lscape} 
\usepackage{multirow}
\usepackage{enumitem,color}

\usepackage[authoryear]{natbib}
\theoremstyle{plain}

\theoremstyle{definition}

\usepackage[]{graphicx}
\chardef\bslash=`\\ 

\newcommand{\com}[1]{\color{black} #1}

\begin{document}
\DOIsuffix{bimj.200100000}
\Volume{52}
\Issue{61}
\Year{2010}
\pagespan{1}{}

\title[Bayesian seamless phase I-II trial design with two stages]{A Bayesian seamless phase I-II trial design with two stages for cancer clinical trials with drug combinations}
\author[Jim\'enez {\it{et al.}}]{Jos\'e L. Jim\'enez\footnote{Corresponding author: {\sf{e-mail: jose\_luis.jimenez@novartis.com}}}\inst{,1,2}} 
\address[\inst{1}]{Biostatistical Sciences and Pharmacometrics, Novartis Pharma A.G., 4056 Basel, Switzerland}
\address[\inst{2}]{Dipartimento di Scienze Matematiche, Politecnico di Torino, Turin, 10129, Italy}
\author[]{Sungjin Kim\inst{3}}
\address[\inst{3}]{Biostatistics and Bioinformatics Research Center, Samuel Oschin Comprehensive Cancer Institute, Los Angeles, CA, 90048, USA}
\author[]{Mourad Tighiouart\inst{3}}
\Receiveddate{zzz} \Reviseddate{zzz} \Accepteddate{zzz} 

\begin{abstract}
The use of drug combinations in clinical trials is increasingly common during the last years since a more favorable therapeutic response may be obtained by combining drugs. In phase I clinical trials, most of the existing methodology recommends a one unique dose combination as ``optimal'', which may result in a subsequent failed phase II clinical trial since other dose combinations may present higher treatment efficacy for the same level of toxicity. We are particularly interested in the setting where it is necessary to wait a few cycles of therapy to observe an efficacy outcome and the phase I and II population of patients are different with respect to treatment efficacy. Under these circumstances, it is common practice to implement two-stage designs where a set of maximum tolerated dose combinations is selected in a first stage, and then studied in a second stage for treatment efficacy. In this article we present a new two-stage design for early phase clinical trials with drug combinations. In the first stage, binary toxicity data is used to guide the dose escalation and set the maximum tolerated dose combinations. In the second stage, we take the set of maximum tolerated dose combinations recommended from the first stage, which remains fixed along the entire second stage, and through adaptive randomization, we allocate subsequent cohorts of patients in dose combinations that are likely to have high posterior median time to progression. The methodology is assessed with extensive simulations and exemplified with a real trial.
\end{abstract}

\maketitle                   






\section{Introduction}

In cancer dose finding clinical trials, the main goal is to identify a safe dose that maximizes the treatment efficacy. In single-agent settings with binary or time to event endpoints, where efficacy is observed relatively fast (e.g. one or two cycles of therapy), one-stage sequential designs where the joint probability of toxicity and efficacy is sequentially updated after each cohort of patients are usually employed (see e.g. \cite{murtaugh1990bivariate, thall1998strategy, braun2002bivariate, ivanova2003new, thall2004dose, chen2015dose, sato2016adaptive} for binary endpoints, and \cite{yuan2009bayesian} for time to event endpoints). Over the years, this methodology has been extended to
 evaluate the mode of action and interaction between combinations of drugs in terms of safety and short-term efficacy endpoints (see e.g. \cite{huang2007parallel, yin2006bayesian} for binary endpoints and \cite{yuan2009bayesian} for time to event endpoints), and proceed in a similar fashion as the methods referenced for single-agent.

However, in settings where efficacy is not ascertained in a short period of time, it is frequent to employ two-stage designs where, a set of maximum tolerated dose combinations is first selected, and then tested for efficacy in a second stage with possibly a different population of patients than the one used in the first stage (see \cite{rogatko2008patient, le2009dose, chen2012methodology}). For drug combination trials, different methodologies for two-stage designs have been proposed for binary efficacy endpoints (see e.g. \cite{shimamura2018two, tighiouart2018two, yuan2011bayesian, zhang2016practical}) but not for time to event endpoints in the second stage of the trial.

One characteristic that most of the previously mentioned methods have in common is that they only recommend a single maximum tolerated dose combination either at the end of the first stage, in two-stage designs, or at the end of the entire trial in cases where toxicity and efficacy are jointly modeled. The reason why this could be a potential problem is that there could be other dose combinations with the same level of toxicity but different level of efficacy.

In this article, we extend the work of \cite{tighiouart2018two} by proposing a new two-stage design for clinical trials with drug combinations with a time to event efficacy endpoint in the second stage and continuous dose combination levels. In the first stage, since the efficacy endpoint is evaluated after three or more cycles of therapy, only binary toxicity data is used to guide the dose escalation. This implies that we have two separate stages in the design we present in this manuscript. The rational for this separation comes from the cisplatin-cabazitaxel trial that motivates this work (see section \ref{ciscabtrial_section}). This trial enrolls patients with advance stage prostate cancer in the first stage whereas in the second stage enrolls only advanced stage patients with visceral metastasis. We may find patients with visceral metastasis in the first stage population, but the number will be low. For this reason, efficacy outcomes (even short term efficacy outcomes) that may come from stage 1 would not be useful for the second stage. Following \cite{tighiouart2017bayesian}, we employ conditional escalation with overdose control (EWOC) to allocate dose combinations to subsequent cohorts of patients and estimate the set of maximum tolerated dose combinations. In the second stage, we propose a Bayesian adaptive algorithm that allocates subsequent cohorts of patients to dose combinations that are likely to have high posterior median time to progression (TTP). To accommodate various functional forms of the median TTP curve, a flexible family of cubic splines is used to model the dose combination - TTP relationship. Adaptive randomization is used sequentially after the disease progression status of each cohort of patients has been resolved and the efficacy curve is updated to minimize the number of patients who are treated at sub-therapeutic dose combinations. At the end of the trial, the dose combination with highest median TTP $\emph{a posteriori}$ is selected and recommended for further phase II or III studies. Our proposal is illustrated by an application to a two-stage dose finding clinical trial that combines cisplatin and cabazitaxel in patients with advanced prostate cancer. A recent phase I clinical trial \cite{lockhart2014phase} explored three dose combinations of these two agents using a `3+3' design. Using a methodology that has been highly criticized (see \cite{paoletti2015statistical}), this trial declared the dose combination 15/75 $mg/m^2$ of cabazitaxel and cisplatin to be the maximum tolerated dose combination even though no dose limiting toxicity (DLT) was observed at this dose combination in the phase I portion of the study. Even though results from this study will be used as prior information, an entirely new dose escalation is implemented in the first stage of our proposal as we hypothesize that a series of tolerable and active dose combinations exist. 

The rest of the manuscript is organized as follows. In sections 2 and 3, we describe the first and second stages of the proposed two-stage design. In section 4, we illustrate the methodology with the two-stage drug combination trial of cisplatin and cabazitaxel in patients with prostate cancer with visceral metastasis where time to progression is a secondary endpoint. The goal in this trial is to find a tolerable dose combination with highest median TTP. A discussion of the approach and final remarks are included in Section 5.


\section{Stage 1}
\label{stage1_section}

\subsection{Model}

Following \cite{tighiouart2017bayesian}, consider the generic form of a dose-toxicity model 

\begin{equation}
\label{modelprobdlt}
P(T=1 | x,y) = F(\eta_0 + \eta_1 x + \eta_2 y + \eta_3 x y),
\end{equation}where $T = 1$ represents an observed toxicity at the dose combination ($x,y$), $T=0$ otherwise, $x \in [X_{\min}, X_{\max}]$ is the dose level of agent $A$, $y \in [Y_{\min}, Y_{\max}]$ is the dose level of agent $B$ and $F(.)$ is a predefined cumulative distribution function. We assume that the dose combinations are continuous and standardized to be in the interval [0,1], the interaction parameter $\eta_3 > 0$, and $\eta_1, \eta_2 > 0$ in order to guarantee that the probability of toxicity increases with the dose of any agent when the other one is held constant. Mind however that $P(T = 1 | x = 0, y = 0) > 0$. Hence, the standardization of $x$ and $y$ does not refer to the probability of toxicity in the absence of the two drugs, just to the dose level of each drug.

The maximum tolerated dose combination is defined as any dose combination $(x^*,y^*)$ such that 

\begin{equation}
P(T=1 | x^*,y^*) = \theta,
\end{equation}where $\theta$ is the target probability of DLT.

As described in \cite{tighiouart2017bayesian}, we reparameterize equation \eqref{modelprobdlt} in terms of parameters that clinicians can easily interpret. One way is to use $\rho_{10}$, the probability of toxicity when the levels of drugs $A$ and $B$ are 1 and 0, respectively, $\rho_{01}$, the probability of toxicity when the levels of drugs $A$ and $B$ are 0 and 1, respectively, $\rho_{00}$, the probability of toxicity when the levels of drugs $A$ and $B$ are both 0. It is possible to show that maximum tolerated dose combination set takes the form

\begin{equation}
\label{mtdcurve}
\begin{split}
& C = \biggl \{ (x^*,y^*): y^* = \biggl [ (F^{-1}(\theta) - F^{-1}(\rho_{00})) - (F^{-1}(\rho_{10}) - F^{-1}(\rho_{00})) x^* \biggr ] \\
& \div  \biggl [ (F^{-1}(\rho_{01}) - F^{-1}(\rho_{00})) + \eta_3 x^* \biggr ] \biggr \}. 
\end{split}
\end{equation}

We assume that $\rho_{10}, \rho_{01}$ and $\eta_3$ are independent \emph{a priori} with $\rho_{01} \sim \mbox{beta}(a_1,b_1)$, $\rho_{10} \sim \mbox{beta}(a_2,b_2)$, and conditional on $(\rho_{01}, \rho_{10})$, $\rho_{00} / \min (\rho_{01}, \rho_{10}) \sim \mbox{beta}(a_3,b_3)$. The prior distribution on the interaction parameter $\eta_3$ is a gamma distribution with mean $a/b$ and variance $a/b^2$. Let $D_n = \{(x_i, y_i, t_i)\}$, where $i=1,\dots, n,$ be the data gathered after enrolling $n$ patients. The posterior distribution of the model parameters is 

\begin{equation}
\begin{split}
&\pi(\rho_{00}, \rho_{10}, \rho_{01}, \eta_3) \propto \prod_{i=1}^{n} G((\rho_{00}, \rho_{10}, \rho_{01}, \eta_3; x_i, y_i))^{t_i} \\
& \times (1 - G(\rho_{00}, \rho_{10}, \rho_{01}, \eta_3; x_i, y_i))^{1-t_i} \\
& \times \pi(\rho_{01}) \pi(\rho_{10}) \pi(\rho_{00} | \rho_{01},\rho_{10}) \pi(\eta),
\end{split}
\end{equation}where

\begin{equation}
\begin{split}
&G(\rho_{00}, \rho_{10}, \rho_{01}, \eta_3; x_i, y_i) = F(F^{-1}(\rho_{00}) + (F^{-1}(\rho_{10}) - \\
& F^{-1}(\rho_{00})) x_i  + (F^{-1}(\rho_{01}) - F^{-1}(\rho_{00})) y_i + \eta_3 x_i y_i).
\end{split}
\end{equation}

Note that the operating characteristics of this stage are evaluated using informative prior distributions (see \cite{tighiouart2018two}).

\subsection{Trial Design}

Dose escalation / de-escalation proceeds using the same algorithm described in \cite{tighiouart2017bayesian}. It is based on EWOC where, after each cohort of enrolled patients, the posterior probability of overdosing the next cohort of patients is bounded by a feasibility bound $\alpha$, see e.g.  \cite{babb1998cancer, tighiouart2005flexible, tighiouart2010dose, tighiouart2012number, shi2013escalation}. In a cohort with two patients, the first one would receive a new dose of agent $A$ given that the dose $y$ of agent $B$ that was previously assigned. The other patient would receive a new dose of agent $B$ given that dose $x$ of agent $A$ was previously assigned. Using EWOC, these new doses are at the $\alpha$-th percentile of the conditional posterior distribution of the maximum tolerated dose combinations. In this design, $\alpha$ increases from 0.25 up to 0.5. The fixed increments of $\alpha$ are set so that $\alpha$ is equal to 0.5 when half of the sample size at stage 1 is enrolled (see \cite{wheeler2017toxicity}). The algorithm continues until the maximum sample size is reached or until the trial is stopped for safety. A more detailed description of both the algorithm and the stopping rule for safety can be found in section S1 of the supplementary material.

Mind that comparisons with other dose escalation approaches similar to EWOC are out of the scope of this article. In case the reader is interested in knowing how the EWOC performs with respect to other approaches see e.g. \cite{tighiouart2017bayesian}

\subsection{Design Operating Characteristics}

At the end of the trial, the estimated maximum tolerated dose combination (MTD) curve is estimated as 

\begin{equation}
\label{estimatedMTDcurve}
\begin{split}
& C_{\tiny \mbox{est}} = \biggl \{ (x^*,y^*): y^* = \biggl [ (F^{-1}(\theta) - F^{-1}(\widehat{\rho}_{00})) - (F^{-1}(\widehat{\rho}_{10}) - F^{-1}(\widehat{\rho}_{00})) x^* \biggr ] \div \\
&  \biggl [ (F^{-1}(\widehat{\rho}_{01}) - F^{-1}(\widehat{\rho}_{00})) \widehat{\eta}_3 x^* \biggr ] \biggr \},
\end{split}
\end{equation}where $\widehat{\rho}_{00}, \widehat{\rho}_{10}, \widehat{\rho}_{01}, \widehat{\eta}$ are the posterior medians given the data $D_n$. Uncertainty about the estimated MTD curve is evaluated through the pointwise average bias and percent of selection as described in \cite{tighiouart2017bayesian}. More precisely, let $C_{\tiny \mbox{est}, i}$ be the estimated MTD curve, and $C$ be the true MTD curve, where $i = 1, \dots, m$ is the number of simulated trials. For every point $(x,y) \in C$, let

\begin{equation}
d_{(x,y)}^{(i)} = \mbox{sign}(y' - y) \min_{(x^*,y^*):(x^*,y^*) \in C} \{ (x - x^*)^2 + (y - y^*)^2 \}^{1/2},
\end{equation}where $y'$ is such that $(x,y') \in C_{\tiny \mbox{est}, i}$. This is the minimum relative distance of the point $(x,y)$ of $C$ to $C_{\tiny \mbox{est}, i}$. Let,

\begin{equation}
\label{equationbias}
d_{(x,y)} = \frac{1}{m} \sum_{i=1}^{m} d_{(x,y)}^{(i)}.
\end{equation}

Equation \eqref{equationbias} can be viewed as the pointwise average bias when estimating the maximum tolerated dose combination. Now let $\Delta (x,y)$ be the euclidean distance between the minimum dose combination $(0,0)$ and the point $(x,y)$ on the MTD curve. Also let $0 < p < 1$, and 

\begin{equation}
\label{equationpercentselection}
P_{(x,y)} = \frac{1}{m} \sum_{i=1}^{m} I \{|d_{(x,y)}^{(i)}| \leq p \Delta (x,y) \}.
\end{equation}

Equation \eqref{equationpercentselection} represents the pointwise percentage of trial for which the minimum distance of the point $(x,y)$ on $C$ to $C_{\tiny \mbox{est}, i}$ is no more than 100p\% of the true maximum tolerated dose combination. In other words, this statistic is equivalent to drawing a circle with radius $p\Delta (x,y)$ and calculating the percentage of trials with $C_{\tiny \mbox{est}, i}$ falling inside the circle, where $p$ is a tolerance parameter. In this article we present the operating characteristics of the first stage of the design in the context of the cisplatin-cabazitaxel trial (see section \ref{ciscabtrial_section}).


\section{Stage 2}
\label{stage2_section}

\subsection{Model}
\label{stage2_model_section}

{\com 
In this section we introduce the model to estimate the dose combination with highest TTP. In this manuscript we use a model with high flexibility in a drug combinations space similar to the one used by \cite{tighiouart2018two}. However, working on two dimensions with a high level of flexibility would result in a model with too many parameters. We propose to reduce the dimension of the problem by projecting each dose combination $(x,y) \in C_{\tiny \mbox{est}}$ onto $z$, which for convenience is also standardized. Mind that along the manuscript we refer to $z$ as ``a dose combination'' and not just as ``a dose'' because its value is in reality a projection of a dose combination. This process is illustrated in section S4 of the supplementary material with an example of how to from $(x,y) \in C_{\tiny \mbox{est}}$ to $z \in [0,1]$ and from $z \in [0,1]$ to $(x,y) \in C_{\tiny \mbox{est}}$. 
}

Mind that the probability of observing efficacy at $z = 0$ is greater than 0. The standardization of $z$ does not refer to the probability of efficacy.

We model the time to progression as a Weibull distribution with probability density function

\begin{equation}
\label{weibullpdf}
f(t;z) = \frac{k}{\lambda (z; {\boldsymbol \psi})} \left ( \frac{t}{\lambda (z; {\boldsymbol \psi})} \right )^{k-1} \exp \left ( - \frac{t}{\lambda (z; {\boldsymbol \psi})} \right )^k,
\end{equation}where $\lambda > 0$ is the shape parameter and $k>0$ is the scale parameter.

The median TTP is

\begin{equation}
\label{median}
\mbox{Med}(z) = \lambda (z; {\boldsymbol \psi}) (\log 2)^{\frac{1}{k}}.
\end{equation} 

A flexible way of modeling the median TTP along the MTD curve is through the use of the cubic spline function

\begin{equation}
\label{shape_model_eq}
\lambda (z; {\boldsymbol \psi}) = \exp \left ( \beta_0 + \beta_1 z + \beta_2 z^2 + \sum_{j=3}^{5} \beta_j (z - \phi_j)_+^3 \right ),
\end{equation}where ${\boldsymbol \psi} = ({\boldsymbol \beta}, {\boldsymbol \phi})$, with ${\boldsymbol \beta} = (\beta_0, \dots, \beta_5)$ and ${\boldsymbol \phi} = (\phi_3, \dots, \phi_5)$, being $\phi_3 = 0$. Let $D_n = \{ (z_i, t_i, \delta_i), i = 1 \dots, n \}$ be the data after enrolling $n$ patients in the trial where $t$ represents the TTP or last follow-up, and $\delta$ the censoring status, and let $\pi({\boldsymbol \psi}, k)$ be the joint prior density on the parameter vector ${\boldsymbol \psi}$ and $k$. The posterior distribution is

\begin{equation}
\label{posterior}
\begin{split}
&\pi({\boldsymbol \psi, k} | D_m) \propto \pi({\boldsymbol \psi}, k) \prod_{i = 1}^{n} \left [ \frac{k}{\lambda (z_i; {\boldsymbol \psi})} \left ( \frac{t_i}{\lambda (z_i; {\boldsymbol \psi})} \right )^{k-1} \right ]^{\delta_i} \times \exp \left ( - \frac{t_i}{\lambda (z_i; {\boldsymbol \psi})} \right )^k.
\end{split}
\end{equation}

We believe the model presented here, due to the cubic spline, allows for great flexibility in terms of identifying the region of dose combinations $z$ with highest TTP. One possible concern this model could raise is the that it has a fairly large number of parameters and a relatively low sample size (30 patients in stage 2). However, these sample sizes are not uncommon in the dose finding literature (see e.g., \cite{tighiouart2018two, wages2014phase}).

Let Med$_z$ be the median TTP at dose combination $z$ and let Med$_0$ be the  median TTP of the standard of care treatment. We propose an adaptive design in order to test the hypothesis

\begin{equation}
\label{eq_hypotheses}
\begin{split}
& \mbox{H}_0: \mbox{Med}_z \leq \mbox{Med}_0 \mbox{ for all } z
\quad 
\mbox{vs.} \\
&\mbox{H}_1: \mbox{Med}_z > \mbox{Med}_0  \mbox{ for some } z.
\end{split}
\end{equation}

It is important to keep in mind that the reason why we can use a model with a fairly large number of parameters is because we work in a continuous dose space. In a discrete dose space, it is not common to test so many dose combinations. Also, a model with a large number of parameters would most likely be non-identifiable, even with large sample sizes. Using the methodology we propose in this manuscript would require to modify \eqref{shape_model_eq}. The use of continuous dose combinations is not uncommon in dose finding studies as we describe in section \ref{ciscabtrial_section} since the drugs are administered intravenously and this allows to administer any drug concentration we need. However, for a larger confirmatory clinical trial, continuous doses are not practical and pills are usually manufactured.

\subsection{Trial Design}

This stage of the trial makes use of response-adaptive randomization to decide in which dose combinations cohorts of patients are allocated. 

We first treat $n_1$ patients at dose combinations in doses equally spaced along the estimated MTD curve $C_{\tiny \mbox{est}}$. Then, we generate $n_2$ dose combinations from the standardized density function \eqref{median} using the posterior distribution of the model parameters that already accounts for the observed efficacy outcomes. The rational for this approach is based on the rejection-sampling principle, which can be used to generate observations from a target distribution (in our case \eqref{median}). Hence, if we generate data from \eqref{median}, we will be allocating patients to dose combinations that are more likely to have higher TTP according to the current estimation of \eqref{median} (i.e., the shape of \eqref{median} will be updated as patients enroll). We will keep allocating patients in cohorts of size $n_2$ following the rejection-sampling principle until the maximum sample size is reached (or the trial is stopped due to futility or safety).

Mind that, allocating the first cohort of $n_1$ patients equally spaced has the sole purpose of gathering efficacy information spaced along the MTD curve since we assume no prior information regarding which region of the MTD curve has higher TTP. This of course, implies allocating patients in high/low pair of dose combinations (i.e., the extremes of the MTD curve), which unless we are working with biologics, would not be optimal \emph{a priori} under the monotonicity assumption. However, in a drug combination setting, the relationship between efficacy and the dose combinations is unknown, and will very much depend on the type of drug we are combining. For this reasons, we believe it is a more suitable approach to explore efficacy along the MTD curve even if it implies allocating patients to high/low pairs of dose combinations. Moreover, even though stage 2 does not use toxicity data to find the optimal dose combination, it contains a stopping rule for safety using a Bayesian continuous monitoring approach (see e.g., \cite{yu2016group}). Overall, the stage 2 can be viewed as an extension of a response-adaptive randomization procedure with a finite number of doses to a response-adaptive randomization procedure with an infinite number of dose combinations.

A more detailed description of this algorithm, including the stopping rules, is presented in the supplementary material.

\subsection{Design Operating Characteristics}
\label{operating_characteristics_ch}

We assess the operating characteristics of the proposed design by assuming that $\lambda(z;{\boldsymbol \psi})$ is a cubic spline with two knots placed between 0 and 1. This class of modeling is very flexible and is able to adapt to scenarios where the median TTP is either constant or skewed toward one of the edges. Vague priors are placed on the model parameters by assuming ${\boldsymbol \beta} \sim N({\boldsymbol \mu}, \sigma^2 {\boldsymbol I}_6)$, where ${\boldsymbol \mu= \{0,0,0,0,0,0 \}}$ and $\sigma^2 = 100$, $(\phi_4, \phi_5) \sim \mbox{Unif} \{(u,v): 0 \leq u < v \leq 1\}$, and $k \sim \mbox{Unif}(10^{-100},10)$. Note that the hyper-parameters of the prior distributions are always the same regardless the value of Med$_0$. {\com In Table \ref{table_induced_prior_medianTTP} we present, for different dose combinations $z$, the induced median TTP with a 90\% CI. The purpose of this table is show the reader the estimated median TTP with a 90\% credible intervals (CI) if we only use the prior distributions. In case we would have an idea \emph{a priori} of the median TTP value, the prior distributions should be chosen so that the induced median TTP is close to that value with 90\% CI not too wide. In our case, since we assume no prior knowledge about the median TTP, we see a very wide 90\% CI, which means that the posterior distribution will be driven by the data we observe during the trial rather than by the prior distributions.}

\begin{table}[t]
\centering
\caption{Induced prior median TTP with 90\% credible intervals (CI) for different dose combinations $z$.}
\begin{tabular}{cccc} \hline
\multicolumn{4}{c}{Induced prior median TTP + 90\% CI}                            \\ \hline
\multirow{2}{*}{Dose combination $z$} & \multicolumn{3}{c}{Quantile} \\
 & 5\% & 50\% & 95\% \\ \hline
0.0 & 2.58e-72  & 0.70 & 2.62e+71  \\
0.1 & 1.16e-72  & 0.68 & 6.35e+71  \\
0.2 & 9.66e-74  & 0.65 & 7.58e+72  \\
0.3 & 1.02e-75  & 0.66 & 6.15e+74  \\
0.4 & 8.69e-79  & 0.71 & 5.97e+77  \\
0.5 & 2.75e-83  & 0.72 & 2.01e+82  \\
0.6 & 8.01e-90  & 0.76 & 6.52e+88  \\
0.7 & 3.57e-99  & 0.84 & 1.21e+98  \\
0.8 & 3.49e-112 & 0.81 & 1.11e+111 \\
0.9 & 8.66e-130 & 0.77 & 2.93e+128 \\
1.0 & 8.13e-153 & 0.66 & 2.25e+151 \\ \hline
\end{tabular}
\label{table_induced_prior_medianTTP}
\end{table}

We present the median posterior probability that the median TTP, evaluated at any dose combination $z$, is larger than the null median TTP (Med$_0$) which is defined as $P(\mbox{Med}(z;{\boldsymbol \psi}_i) > \mbox{Med}_0 | D_{n,i})$ for the i-$th$ trial. This median posterior probability is used to obtain the optimal dose combination which is defined as

\begin{equation}
\label{eq_estimated_optimal_dose}
z_i^{\tiny \mbox{opt}} = \mbox{arg max}_v \{ P(\mbox{Med}(v;{\boldsymbol \psi}_i) > \mbox{Med}_0 | D_{n,i}) \}.
\end{equation}

For each scenario favoring the alternative hypothesis, we estimate the Bayesian power, which is defined as

\begin{equation}
\label{power}
\mbox{Power} \approx \frac{1}{M} \sum_{i=1}^{M} I [ \mbox{Max}_z \{ P(\mbox{Med}(z;{\boldsymbol \psi}_i) > \mbox{Med}_0 | D_{n,i}) \} > \delta_u ],
\end{equation}where

\begin{equation}
P(\mbox{Med}(z;{\boldsymbol \psi}_i) > \mbox{Med}_0 | D_{n,i}) \approx \frac{1}{L} \sum_{j=1}^{L} I \left [ \mbox{Med}(z;{\boldsymbol \psi}_{i,j}) > \mbox{Med}_0 \right ]
\end{equation}and ${\boldsymbol \psi}_{i,j}$ is the j-$th$ MCMC sample from the i-$th$ trial.

Mind that for scenarios favoring the null hypothesis, \eqref{power} is the estimated Bayesian type-I error probability.

\section{Application to the cisplatin-cabazitaxel trial}
\label{ciscabtrial_section}

The methodology proposed by \cite{tighiouart2018two} was used to design a two-stage clinical trial of the combination cisplatin and cabazitaxel in patients with prostate cancer with visceral metastasis, where TTP is a secondary endpoint in stage 2. In this section we illustrate the methodology proposed in sections \ref{stage1_section} and \ref{stage2_section} to evaluate the operating characteristics of the clinical trial as a whole. It is important to mention that the first stage of the cisplatin-cabazitaxel trial was motivated by a phase I trial published by \cite{lockhart2014phase} in patients with advanced solid tumors, where the combinations of cisplatin/cabazitaxel 15/75 $mg/m^2$, 20/75 $mg/m^2$ and 25/75 $mg/m^2$ were explored and the maximum tolerated dose combination was established at 15/75 $mg/m^2$. Despite the criticism toward cohort expansion designs (see e.g. \cite{paoletti2015statistical}) a ``3+3'' design was used in this trial. In the first part of this trial, 9 patients were evaluated for safety and no toxicity was observed at 15/75 $mg/m^2$. In the second part, 15 patients were treated at 15/75 $mg/m^2$ and two toxicities were observed. As pointed out by \cite{tighiouart2018two}, on the basis of the results from \cite{lockhart2014phase} and other preliminary efficacy data, it has been hypothesized that there is a series of dose combinations which are tolerable and efficacious in prostate cancer.

The cisplatin-cabazitaxel trial (see \cite{tighiouart2018two}) used a continuous set of doses that range from 10 to 25 mg/$m^2$ for cabazitaxel, and from 50 to 100 mg/$m^2$ for cisplatin, that will be administered intravenously. In a first stage, the cisplatin-cabazitaxel trial trial will enroll 30 patients in order to obtain the MTD curve. This stage of the design proceeds as explained in section \ref{stage2_section}, with a target probability of toxicity $\theta = 0.33$, and a logistic link fuction $F(.)$ in equation \eqref{modelprobdlt}. The starting dose combination for the first cohort of two patients is 15/75 mg/$m^2$, and toxicities are to be resolved within 1 cycle of treatment (3 weeks). Prior distributions are calibrated such that the prior mean probability of toxicity at dose combination 15/75 $mg/m^2$ equals $\theta$ (see \cite{tighiouart2018two}). The operating characteristics of this first stage are obtained by simulating $m=1000$ trial replicates following \cite{tighiouart2017bayesian}.

It is important to acknowledge that toxicity attribution was not taken into considerations.  \cite{jimenez2019cancer} showed that under certain situations, partial attribution of dose limiting toxicity can improve the design operating characteristics. However, given the cytotoxic nature of cisplatin and cabazitazel, most of the toxicities would overlap and the number of toxicities that can be attributed would be very low. As a consequence, we believe this design would not benefit from this approach.

\begin{figure}[t]
\caption{True and estimated MTD curves from scenarios 1-12 of the stage 1. The dashed black line represents the true MTD curve, the solid light orange lines represent the 1000 estimated MTD curves from each simulated trial, and the solid red line present the median of these 1000 estimated MTD curves.}
\centering
\includegraphics[scale=0.85]{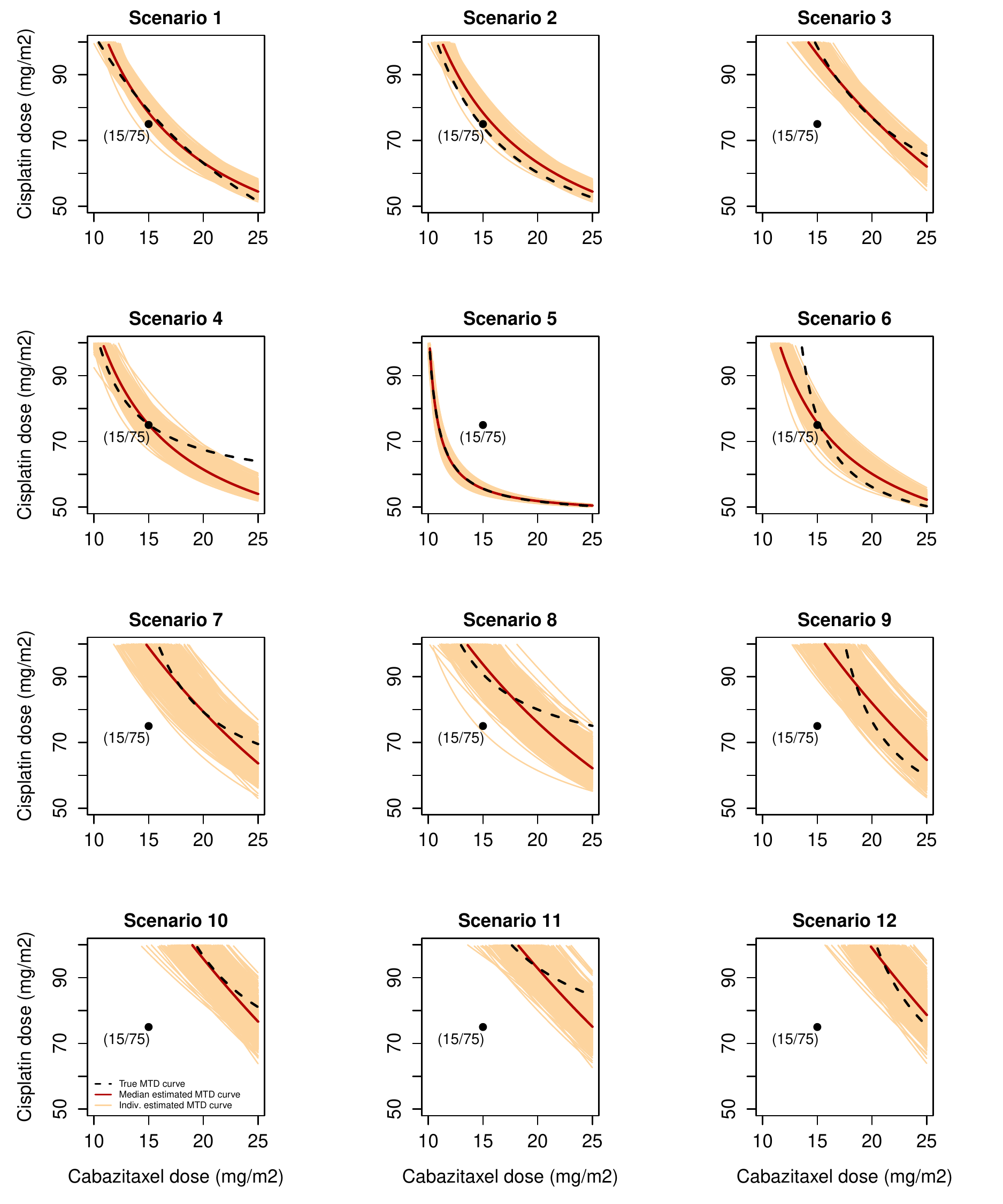}
\label{Figure1_Jimenez2018}
\end{figure}

{\com In Figure 1 we show the true MTD curve (dashed black line), the median estimated MTD curve (solid red line) and the 1000 estimated MTD curves (solid light orange lines) from the 1000 simulated trials in 12 different scenarios. In these scenarios, the true MTD curve goes below (scenarios 2 and 6), above (scenarios 1, 3,7,8,9,10,11,12) and through (scenarios 4 and 6) the dose combination 15/75 mg/$m^2$, also passes through it. We also considered the case where one of the drugs is more toxic than the other. Mind that the reason why there more scenarios where the MTD is placed above the dose combination 15/75 mg/$m^2$ is because this dose combination is quite low considering the entire space of doses and we wanted to be sure we test the design at high dose combinations. Overall, we observe that in all scenarios, the estimated MTD curves is very close to the true MTD curve. The only scenarios where the performance was a bit worse is when one drug is more toxic than the other one. However, even in this cases, a large part of the estimated MTD curve is very close to the true MTD curve.

As previously mentioned, there is prior information that was used to tune the prior distributions in stage 1. To assess the impact of these informative prior distributions, in Figure S11 (see supplementary material) we present the results of the same 12 scenarios used in Figure 1 using vague prior distributions. Of course, less informative priors will translate into higher variability when estimated the median MTD curve, and this is exactly what we see when comparing Figures \ref{Figure1_Jimenez2018} and S11.
}

{\com These results are supported by the pointwise average bias shown in Figure S2 (see supplementary material). In these scenarios, the pointwise average bias fluctuates between -0.2 and 0.1, depending on the scenario. In terms of safety (see Table S2 in the supplementary material), the percent of trials with toxicity rate above $\theta + 0.1$ is below 13\% in all scenarios, except for scenario 5, with average number of toxicities between 27\% and 35\%. Scenario 5 presents a 43\% number of toxicities and a 54.7\% number of trials with DLT rate higher than $\theta + 0.1$. However, these results are normal since the prior distributions of the model parameters were set assuming that at the combination 15/75 $mg/m^2$ the probability of DLT is 0.3333, where in this scenario, the probability of DLT is much higher at this dose combination. We also present results regarding the percentage of correct recommendation. These results are shown in Figure S3 (see supplementary material) and overall we observe that, in the 12 studied scenarios, the percent of correct recommendation is always above 80\% when the design parameter $p=0.2$. When the design parameter $p=0.1$, the performance varies depending on the scenario. In scenarios 1,2,3 and 5, the percent of correct recommendation is above 90\% for all dose combinations. However, for scenarios 4, 6-12 where one drug is more toxic than the other one or the true MTD is clearly above the combination 15/75 $mg/m^2$, the percent of correct recommendations fluctuates, depending on the dose combinations, from 0\% to 100\%. However, in scenarios 4 and 6, the true MTD curves passes through the dose combination 15/75 mg/$m^2$, and at this dose combination the percent of correct recommendation is above 90\%. Again, mind that these results depend on a design parameter $p$ that takes the values 0.1 and 0.2 and that states how strict we are when considering a correct recommendation, being $p=0.2$ less strict than $p=0.1$. This measurement can be understood as a confidence interval. The true parameter values as well as the safety results are shown in Table S2 (see supplementary material). 
}

In the second stage, 30 additional patients are enrolled to identify the dose combinations along the MTD from the first stage, that are likely to have high posterior median TTP. The median TTP of the standard care of treatment, which is necessary to perform the hypothesis testing procedure, is chosen to be 4 months since this is the radiographic median TTP in a placebo arm in a previous phase III trial (see \cite{loriot2017enzalutamide}). We present simulations based on 6 scenarios supporting both the alternative and the null hypotheses. For each scenario favoring the alternative hypothesis, we use effect sizes of 1.5 and 2 months and accrual rates of 1 and 2 patients per month. This way, the only difference between scenarios will be the shape of the median TTP curve, allowing to see the behavior of the design when the optimal dose level is located at different dose combinations $z$. The parameter values used to generate these scenarios are presented in Table S3 (see supplementary material). It is important to note that if during the second stage, the rate of DLT is found to be excessive, the trial is suspended and the protocol may be amended to re-estimate the MTD curve using all toxicity data from the two stages.

The simulations were carried out using the model and prior distributions presented in sections \ref{stage2_model_section} and \ref{operating_characteristics_ch} respectively, with $n_1 = 10$, $n_2 = 5$, $\delta_u = 0.8$ and $\delta_u = 0.9$. 

\begin{figure}[ht!]
\caption{True median TTP curve scenarios under H$_0$ and under H$_1$ for effect sizes (ES) of 1.5 and 2 months (see null and alternative hypotheses in equation \eqref{eq_hypotheses}).}
\centering
\includegraphics[scale=0.7]{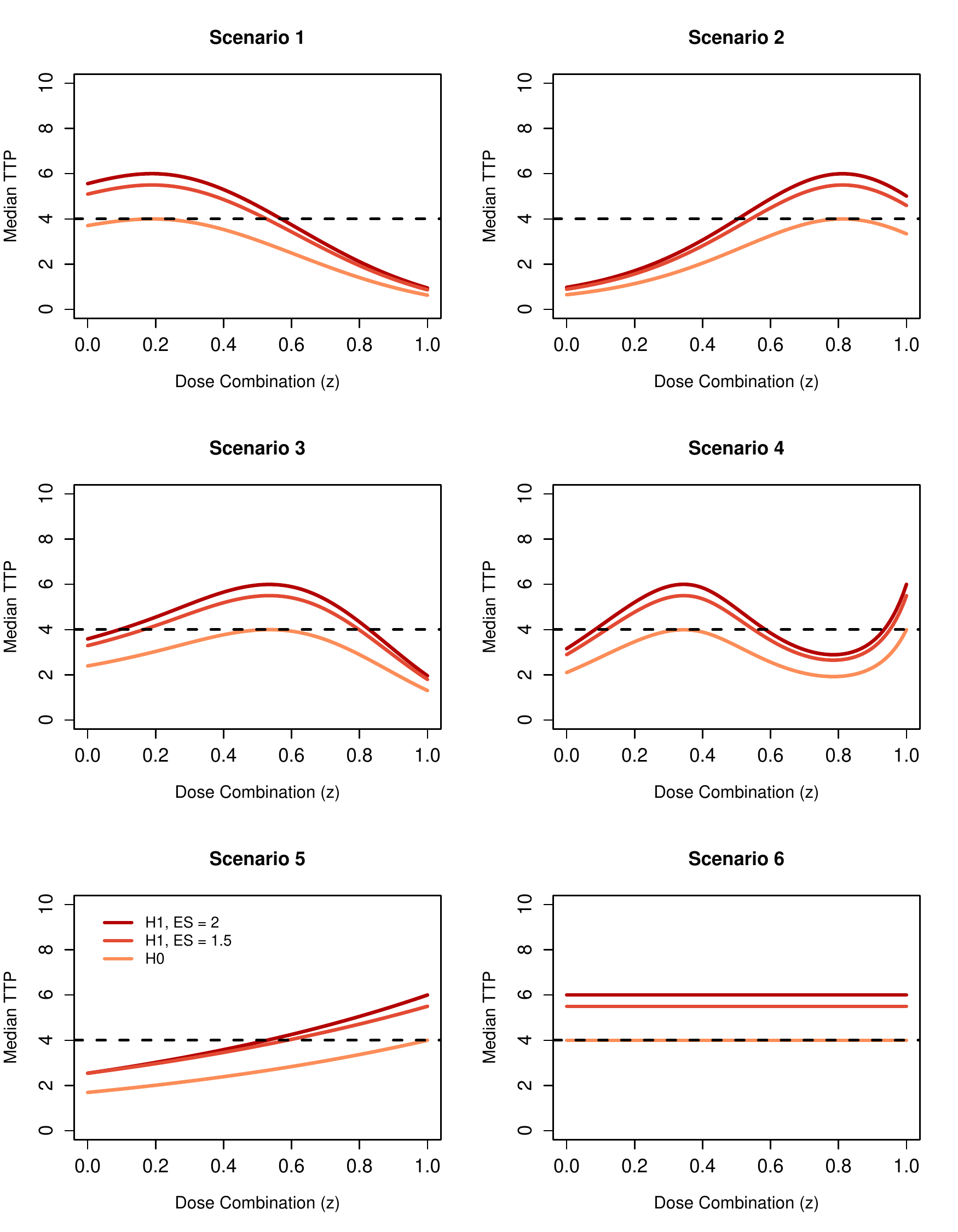}
\label{Figure2_Jimenez2018}
\end{figure}

\begin{figure}[ht!]
\caption{Posterior probability that the median TTP, evaluated at any drug combination $z$, is higher than the null median TTP for effect sizes (ES) of 1.5 and 2 months, accrual rates (AR) of 1 and 2 patients per month, under H$_0$ and H$_1$ (see null and alternative hypotheses in equation \eqref{eq_hypotheses}).}
\centering
\includegraphics[scale=0.7]{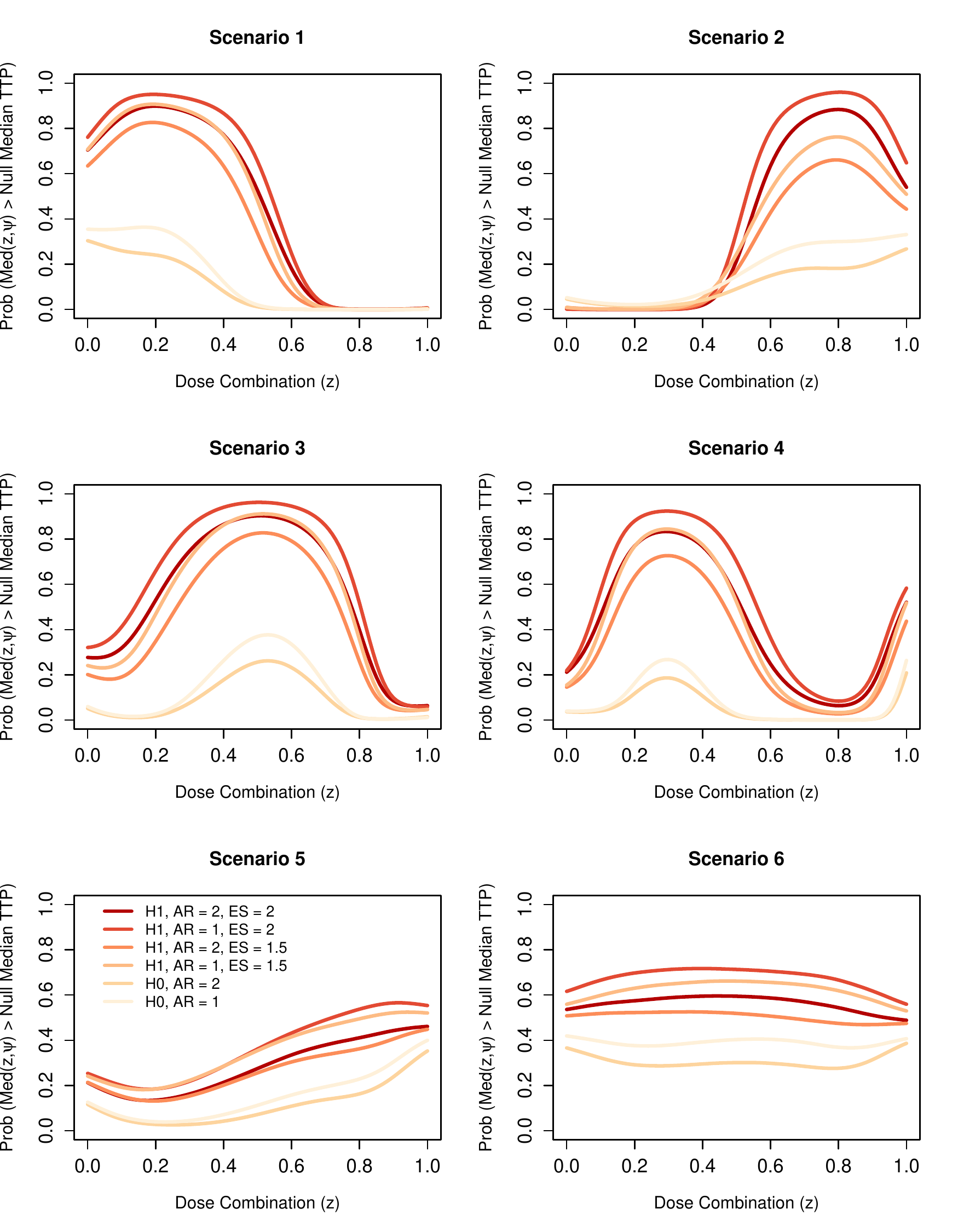}
\label{Figure3_Jimenez2018}
\end{figure}

In Figure \ref{Figure2_Jimenez2018} we present the true median TTP curve of 6 simulated scenarios favoring both the alternative hypothesis with effect sizes of 1.5 month (red line) and 2 (dark red line) months, and the null hypothesis (orange line). In Figure \ref{Figure3_Jimenez2018}, we present the posterior probability that the median TTP, evaluated at any dose combination $z$, is higher than the null median TTP (i.e., $P(\mbox{Med}(z;{\boldsymbol \psi}) > \mbox{Med}_0 | D_n)$). Mind that, after each simulated trial, we obtain a different posterior probability that a dose combination $z$ is higher than the null median TTP. Figure \ref{Figure3_Jimenez2018} presents the median of all these posterior probabilities. Overall, we observe that, regardless the shape of the median TTP, the effect size or the accrual rate, the posterior probability that a dose combination $z$ is higher than the null median TTP reaches its highest values at dose combinations $z$ where the true median TTP is higher. This implies that the estimated optimal dose combination (see equation \ref{eq_estimated_optimal_dose}) is very close to the true optimal dose combination. In the supplementary material, in Figures S5-S10, we present the individual posterior probabilities that a dose combination $z$ is higher than the null median TTP from the 1000 simulated trials as well as their median value for all scenarios, effect sizes and accrual rates.

In Table \ref{table1}, we present the Bayesian power, the probability of the type-I error as well as the probability of type-I + type-II errors for different effect sizes and different accrual rates. With an accrual rate of 1 patient per month, the probability of type-I error remains between 0.104 and 0.384 when $\delta_u = 0.8$ and between 0.0043 and 0.192 when $\delta_u = 0.9$. However, with an accrual rate of 2 patients per month, the probability of type-I error is much smaller overall and it remains between 0.035 and 0.25 when $\delta_u = 0.8$ and between 0.008 and 0.108 when $\delta_u = 0.9$.

\begin{table}[t]
\centering
\caption{Bayesian power, type I error probability and type-I + type-II error probability in scenarios 1-6 with effect sizes of 1.5 and 2 months, and accrual rates of 1 and 2 patients per month.}
\resizebox{\columnwidth}{!}{%
\begin{tabular}{cccccccccccc} \hline
  &   & \multicolumn{2}{c}{\makecell{Power \\ (effect size \\ of 1.5 months)}} & \multicolumn{2}{c}{\makecell{Power \\ (effect size \\ of 2 months)}} & \multicolumn{2}{c}{\makecell{Probability of \\ type-I error}} & \multicolumn{2}{c}{\makecell{Probability of \\ type-I + type-II \\ errors (effect \\ size of 1.5 months)}} & \multicolumn{2}{c}{\makecell{Probability of \\ type-I + type-II \\ errors (effect \\ size of 2 months)}} \\ \hline
  & & \multicolumn{2}{c}{$\delta_u$} & \multicolumn{2}{c}{$\delta_u$} & \multicolumn{2}{c}{$\delta_u$} & \multicolumn{2}{c}{$\delta_u$} & \multicolumn{2}{c}{$\delta_u$} \\ \hline
Scenario &  \makecell{Accrual \\ rate} & 0.8 & 0.9 & 0.8 & 0.9 & 0.8 & 0.9 & 0.8 & 0.9 & 0.8 & 0.9 \\ \hline
1 &      \multirow{6}{*}{1}    & 0.924 & 0.844  & 0.971 & 0.927 & 0.227 & 0.121 & 0.303  & 0.277 & 0.256 & 0.194\\
2 &   & 0.706 & 0.520 & 0.972 & 0.920 & 0.122 & 0.043 & 0.416 &  0.523  & 0.150  & 0.123 \\
3 &    & 0.904 & 0.808 &   0.973 & 0.932  & 0.139 & 0.072 & 0.235 &  0.264  &  0.166 & 0.140 \\
4 &    & 0.796  & 0.646  & 0.931 & 0.846 & 0.104 & 0.044 & 0.308  &  0.398  & 0.173 & 0.198 \\
5 &    & 0.413 & 0.247 & 0.449 & 0.248 & 0.145 & 0.070 & 0.732 & 0.823 & 0.696 & 0.822 \\
6 &    & 0.769 & 0.298 & 0.832 & 0.656 & 0.384 & 0.192 & 0.615 & 0.894 & 0.552 & 0.536\\
\hline
1 &  \multirow{6}{*}{2} & 0.824 & 0.674 & 0.920 & 0.829 & 0.107 & 0.048 & 0.283 & 0.374 & 0.187 & 0.219 \\ 
2 &  & 0.522 & 0.338 & 0.865  & 0.755 & 0.035 & 0.008& 0.513 & 0.670 & 0.170 & 0.253 \\
3 & & 0.759 & 0.598 & 0.896 & 0.790  & 0.068 &0.024 & 0.309 & 0.426 & 0.172 & 0.234\\
4 & & 0.623 & 0.445 & 0.766 &  0.615 & 0.053 & 0.018 & 0.430 & 0.573 & 0.287 & 0.403 \\
5 &    & 0.207 & 0.084 & 0.265 & 0.124 & 0.091 & 0.028 & 0.884 & 0.944 & 0.826 & 0.904 \\
6 &    & 0.547 & 0.298 & 0.627 & 0.394 & 0.250 & 0.108 & 0.703 & 0.810 & 0.623 & 0.714 \\
\hline
\end{tabular}
}
\label{table1}
\end{table}

In terms of power, with an effect size of 1.5 months and an accrual rate of 1 patient per month, we observe that the power remains between 0.706 and 0.924 when $\delta_u = 0.8$ for scenarios 1-4 and 6, and around 0.41 for scenario 5. When $\delta_u = 0.9$ the power ranges between 0.52 and 0.844 for scenarios 1-4 and between 0.247 and 0.298 for scenarios 5-6 . If the effect size increases up to 2 months with an accrual rate of 1 patient per month, the power remains between 0.832 and 0.972 for scenarios 1-4 and 6 when $\delta_u = 0.8$, and between 0.846 and 0.932 for scenarios 1-4, 0.656 for scenario 6 and 0.248 for scenario 5 when $\delta_u = 0.9$. In contrast, if we fix the accrual rate to 2 patient per month, we observe that overall the power decreases considerately. With an effect size of 1.5 months, the power remains between 0.522 and 0.824 for scenario 1-4 and 6, and 0.207 for scenario 5 when $\delta_u = 0.8$, and between 0.298 and 0.674 for scenarios 1-4 and 6, and 0.084 for scenario 5 when $\delta_u = 0.9$. If the effect size increases up to 2 months and maintaining an accrual rate of 2 patients per month, the power remains between 0.766 and 0.92 for scenarios 1-4, 0.627 for scenario 6 and 0.265 for scenario 5 when $\delta_u = 0.8$ and between 0.615 and 0.829 for scenario 1-4, 0.394 for scenario 6 and 0.124 for scenario 5 when $\delta_u = 0.9$.

Because it is difficult to find the right balance between power and type-I error, and since it is not unusual to find probabilities of type-I error between 0.15 - 0.2 in randomized phase II trials, we report the sum of the probabilities of type-I error and type-II errors. In general, a design where the sum of these two probabilities is above 0.3 is not advisable. In our proposal, in scenarios 1-4, with effect size of 1.5 months and an accrual rate of 1 patient per month, the sum of the probabilities of type-I error and type-II error remains between 0.235 and 0.416 when $\delta_u = 0.8$ and between 0.264 and 0.523 when $\delta_u = 0.9$. If the effect size increases up to 2 months and maintaining the same accrual rate, the sum of the probabilities of type-I error and type-II error remains between 0.15 and 0.256 when $\delta_u = 0.8$ and between 0.123 and 0.198 when $\delta_u = 0.9$. If we fix the accrual rate to 2 patient per month, with an effect size of 1.5 months, the sum of the probabilities of type-I error and type-II error remains between 0.283 and 0.513 when $\delta_u = 0.8$ and between 0.374 and 0.67 when $\delta_u = 0.9$. If the effect size increases up to 2 months and maintaining the same accrual rate, the the sum of the probabilities of type-I error and type-II error remains between 0.17 and 0.287 when $\delta_u = 0.8$ and between 0.219 and 0.403 when $\delta_u = 0.9$. In scenarios 1-5, the sum of type I and type II errors is much higher than in other scenarios reaching values above 0.61 with an effect size of 1.5 months and values above 0.53 with an effect size of 2 months. 

If we focus on the sum of the probabilities of type-I and type-II errors for scenarios 1-4, we observe that with an effect size of 1.5 months we observe a lot of values above our 0.3 threshold regardless the accrual rate, which is normal since the original design's primary endpoint was not the TTP median and it is not sufficiently powered for this effect size. In contrast, with an effect size of of 2 months, we observe that if $\delta_u=0.8$ all the values are below our 0.3 threshold regardless the accrual rate and if $\delta_u=0.9$, only one scenario with an accrual rate of 2 patients per month has a value above the threshold. Regarding scenarios 5-6, the sum of the type I and type II errors is too high and hence we would not advice the use of our design in cases where there is prior belief that the median TTP curve shape is similar to the shapes studied in these two scenarios.

In Table \ref{table2}, we present the probability of early stopping and average sample size at the moment of stopping in scenarios favoring the null hypothesis. Overall, we observe that an accrual rate of 2 patients per month produces a slight increase in the probability of early stopping and a decrease between 1 and 2 patients in the average sample size at the moment of stopping with respect to using an accrual rate of 1 patient per month.

\begin{table}[t]
\centering
\caption{Probability of early stopping under the null hypothesis in scenarios 1-6 with accrual rates of 1 and 2 patients per month.}
\begin{tabular}{ccccccccc} \hline
    &     & \multicolumn{3}{c}{\makecell{Probability of \\ early stopping}} & & \multicolumn{3}{c}{Average sample size} \\ \hline
    &     & \multicolumn{3}{c}{$\delta_0$} & & \multicolumn{3}{c}{$\delta_0$} \\ \hline 
Scenario &  \makecell{Accrual \\ rate} & 0.10     & 0.15     & 0.20  &   & 0.10        & 0.15        & 0.20        \\ \hline
1  &  \multirow{6}{*}{1}  &     0.164   &   0.266   & 0.355  &   & 19.94  & 18.14 & 16.92 \\
 2  &   & 0.121 & 0.252 & 0.390 &   & 17.44 & 15.41 & 13.76 \\
 3  &   & 0.234 & 0.358 & 0.506 &   & 20.17 & 17.67 & 16.13 \\
 4  &   & 0.264 & 0.417 & 0.554 &  & 19.79 & 17.75 & 16.17 \\
 5  &   & 0.095 & 0.212 & 0.336 &  & 18.53 & 16.79 & 15.53 \\
 6  &   & 0.016 & 0.034 & 0.062 &  & 16.25 & 14.69 & 12.81 \\\hline
1  &  \multirow{6}{*}{2}  & 0.307 & 0.457 & 0.576 &   & 18.55 & 16.95 & 15.99 \\
 2  &   & 0.270 & 0.452 & 0.611 &   & 15.64 & 13.63 & 12.20 \\
 3  &   & 0.410 & 0.611 & 0.740 &   & 19.02 & 16.94 & 15.01 \\
 4  &   & 0.421 & 0.590 & 0.731 &  & 18.41 & 16.44 & 14.67 \\
 5  &   & 0.213 & 0.376 & 0.529 &  & 16.15 & 13.94 & 12.77 \\
 6  &   & 0.040 & 0.095 & 0.161 &  & 12.13 & 11.38 & 11.13 \\\hline
\end{tabular}
\label{table2}
\end{table}

Even though it is not listed in the operating characteristics, in Figure S4 (see supplementary material) we show the dose allocation distribution in the 6 studied scenarios with the different effect sizes and accrual rates. In scenario 1, we correctly allocate more than 71\% of the patients in doses that are above the TTP of the standard treatment of care. In scenarios 2, 3, 4 and 5 we correctly allocate more than 65\%, 77\%, 60\% and 52\% of the patients respectively in dose above the TTP of the standard treatment of care. In scenario 6 all dose combinations are above the TTP of the standard treatment of care  and hence 100\% are correctly allocated. Note that from these distributions we excluded the first $n_1$ doses which are automatically allocated in doses equally spaced along the maximum tolerated dose combination.

All the simulations presented so far have been done assuming a null median TTP of 4 months since, as previously explained, this is the radiographic median TTP in a placebo arm from a previous phase III trial in patients with prostate cancer with visceral metastasis. However, the methodology proposed in this article may be used in another setting where the null median TTP is higher than 4 months. To illustrate the performance of our proposal assuming a higher null median TTP, we run scenarios 1 and 2 with an effect size equal to 2 months, an accrual rate of 2 and 3 patients per month, and a the null median TTP equal to 8 months (see Table S1 in the supplementary material). Overall, we observe values of power, type-I error and sum of type-I and type-II errors that are consistent with the values presented in Table \ref{table1}.

\section{Conclusions}

In this paper we propose a Bayesian two-stage design for cancer clinical trials using drug combinations with continuous dose levels. We are motivated by the ``cisplatin-cabazitaxel trial'', a two stage design that combines cisplatin and cabazitaxel in patients with prostate cancer with visceral metastasis, where TTP is a secondary endpoint in stage 2. TTP is defined as time to disease progression or death. Since the sample size for stage 2 is relatively small and we do not expect many death by 6 months or so, the use of a competing risk model to estimate the cumulative incidence of disease progression may not be different from the marginal estimate of time to disease progression.

In the first stage, we studied the operating characteristics under six scenarios. In four of the scenarios, the true MTD passes through the dose combination 15/75 mg$/m^2$, a dose previously identified as an MTD, whereas in the other two scenarios the true MTD is above and below this combination. We found that this stage of the trial is safe and has good operating characteristics in terms of pointwise bias and percent selection. Note that the operating characteristics of this stage were evaluated using informative prior distributions as discussed in section \ref{ciscabtrial_section}. With respect to the second stage, we evaluated the performance of the design under six scenarios including linear and non-linear median TTP curve shapes. We conclude that our design has overall good operating characteristics with accrual rates that are considered realistic in practice. However, we observed superior performance in scenarios with non-linear dose-efficacy relationships. Hence, if there is prior belief that the median TTP curve shape is linear, we advise the use of a different model than the one with the cubic spline proposed in equation \eqref{median}. This is a limitation of our model and is a consequence of a small sample size in stage 2. Note that these operating characteristics were evaluated under vague prior distributions for the model parameters and no efficacy profiles of single agent trials were used \emph{a priori}.

{\com

One characteristic of this design is that, stage 1 does not make use of efficacy outcomes. Hence, we have two stages that are disconnected in the sense that stage 1 is driven only by toxicity data and stage 2 is driven only by efficacy data (although toxicity is monitored in stage 2 to stop the trial if necessary). As previously described, the rational for this separation comes from the fact that patients in the two stages have different disease profiles; in stage 1, patients have advance stage prostate cancer whereas in stage 2, patients have advanced stage prostate cancer but with visceral metastasis. According to the clinician who designed the trial, the efficacy outcomes of these two different patient populations are expected to be very different. In addition, a Bayesian continuous monitoring of the rate of toxicity in stage 2 is also carried out as discussed in \cite{tighiouart2018two} so that the trial will stop early if there is evidence of an excessive rate of toxicity and the protocol may be amended to re-estimate the MTD curve using all toxicity data from both stages.

Another issue with this methodology is that uncertainty of the estimated MTD in stage 1 is not taken into account in stage 2 of the design. This implies that the MTD is not updated during the second stage, which is a limitation since patients in stage 2 may come from a different population with respect to patients in the first stage. As pointed out by \cite{tighiouart2018two}, an alternative design for this particular paper would account for first-, second- and third-cycle toxicity in addition to efficacy outcome at each cycle. This is subject to future research but such model is unlikely to improve trial outcome in this setting due to the small sample size in stage 2 and increased number of parameters. We also note that no response adaptive randomization was used for the first cohort of patients in stage 2. The first cohort of patients in stage 2 are allocated homogeneously along the MTD curve recommended from stage 1. This would make sense if we work with biologics but it may not be ethical for two cytotoxic agents where the most efficacious dose combination is expected to be located in the middle of the MTD curve. However, this assumes a certain level of activity and synergism between the two drugs and we believe that having efficacy data from the entire MTD curve initially will provide a better estimate of the median TTP curve function.

Last, this work generates some open questions that would require further research. For example, alternatives to the TTP efficacy curves using  cubic splines may be investigated when simpler patterns of this curve are expected. Another point that will be tackled in the future is a comparison with other existing methods even if there is a very limited number of approaches that are similar to ours. There exists however different approaches for stage 1 alone as we mentioned along this manuscript. Hence, it would be interesting to compare the performance of all the existing dose escalation methods under different scenarios as this would help to know which approaches work better under which circumstances and to justify the design that is selected. Finally, an extension of this two-stage model that varies the MTD curve in stage 2 as new toxicity data are accumulated during this stage is under consideration.
}
\begin{acknowledgement}
We thank the three anonymous reviewers and the Associate Editor for their valuable feedback that helped improve the quality of this article. This project has received funding from the European Union's Horizon 2020 research and innovation programme under the Marie Sklodowska-Curie grant agreement No 633567 (J.J.), the National Institute of Health Grant Number 1R01CA188480-01A1 (M.T.), the National Center for Research Resources, Grant UL1RR033176, and is now at the National Center for Advancing Translational Sciences, Grant UL1TR000124 (M.T.), and 2 P01 CA098912 (M.T.).
\end{acknowledgement}
\vspace*{1pc}

\noindent {\bf{Conflict of Interest}}

\noindent {\it{The authors have declared no conflict of interest.}}

\vspace*{1pc}

\noindent {\bf{Disclaimer}}

\noindent {\it{The views and opinions expressed in this article are those of the author (J.J.) and do not necessarily reflect the official policy or position of Novartis Pharma A.G.}}


\bibliographystyle{biometrical}
\bibliography{references.bib}

\begin{thebibliography}{30}
\providecommand{\natexlab}[1]{#1}
\providecommand{\url}[1]{\texttt{#1}}
\providecommand{\urlprefix}{URL }

\bibitem[{Babb \emph{et~al.}(1998)Babb, Rogatko, and Zacks}]{babb1998cancer}
Babb, J., Rogatko, A., and Zacks, S. (1998).
\newblock Cancer phase I clinical trials: efficient dose escalation with
  overdose control.
\newblock \emph{Statistics in medicine} \textbf{17}, 1103--1120.

\bibitem[{Braun(2002)}]{braun2002bivariate}
Braun, T.~M. (2002).
\newblock The bivariate continual reassessment method: extending the CRM to
  phase I trials of two competing outcomes.
\newblock \emph{Controlled clinical trials} \textbf{23}, 240--256.

\bibitem[{Chen \emph{et~al.}(2015)Chen, Yuan, Li, Kutner, Owonikoko
  \emph{et~al.}}]{chen2015dose}
Chen, Z., Yuan, Y., Li, Z., Kutner, M., Owonikoko, T., \emph{et~al.} (2015).
\newblock Dose escalation with over-dose and under-dose controls in Phase I/II
  clinical trials.
\newblock \emph{Contemporary clinical trials} \textbf{43}, 133--141.

\bibitem[{Chen \emph{et~al.}(2012)Chen, Zhao, Cui, and
  Kowalski}]{chen2012methodology}
Chen, Z., Zhao, Y., Cui, Y., and Kowalski, J. (2012).
\newblock Methodology and application of adaptive and sequential approaches in
  contemporary clinical trials.
\newblock \emph{Journal of Probability and Statistics} \textbf{2012}.

\bibitem[{Huang \emph{et~al.}(2007)Huang, Biswas, Oki, Issa, and
  Berry}]{huang2007parallel}
Huang, X., Biswas, S., Oki, Y., Issa, J.-P., and Berry, D.~A. (2007).
\newblock A parallel phase I/II clinical trial design for combination
  therapies.
\newblock \emph{Biometrics} \textbf{63}, 429--436.

\bibitem[{Ivanova(2003)}]{ivanova2003new}
Ivanova, A. (2003).
\newblock A New Dose-Finding Design for Bivariate Outcomes.
\newblock \emph{Biometrics} \textbf{59}, 1001--1007.

\bibitem[{Jimenez \emph{et~al.}(2019)Jimenez, Tighiouart, and
  Gasparini}]{jimenez2019cancer}
Jimenez, J.~L., Tighiouart, M., and Gasparini, M. (2019).
\newblock Cancer phase I trial design using drug combinations when a fraction
  of dose limiting toxicities is attributable to one or more agents.
\newblock \emph{Biometrical Journal} \textbf{61}, 319--332.

\bibitem[{Le~Tourneau \emph{et~al.}(2009)Le~Tourneau, Lee, and
  Siu}]{le2009dose}
Le~Tourneau, C., Lee, J.~J., and Siu, L.~L. (2009).
\newblock Dose escalation methods in phase I cancer clinical trials.
\newblock \emph{JNCI: Journal of the National Cancer Institute} \textbf{101},
  708--720.

\bibitem[{Lockhart \emph{et~al.}(2014)Lockhart, Sundaram, Sarantopoulos, Mita,
  Wang-Gillam \emph{et~al.}}]{lockhart2014phase}
Lockhart, A.~C., Sundaram, S., Sarantopoulos, J., Mita, M.~M., Wang-Gillam, A.,
  \emph{et~al.} (2014).
\newblock Phase I dose-escalation study of cabazitaxel administered in
  combination with cisplatin in patients with advanced solid tumors.
\newblock \emph{Investigational new drugs} \textbf{32}, 1236--1245.

\bibitem[{Loriot \emph{et~al.}(2017)Loriot, Fizazi, de~Bono, Forer, Hirmand
  \emph{et~al.}}]{loriot2017enzalutamide}
Loriot, Y., Fizazi, K., de~Bono, J.~S., Forer, D., Hirmand, M., \emph{et~al.}
  (2017).
\newblock Enzalutamide in castration-resistant prostate cancer patients with
  visceral disease in the liver and/or lung: Outcomes from the randomized
  controlled phase 3 AFFIRM trial.
\newblock \emph{Cancer} \textbf{123}, 253--262.

\bibitem[{Murtaugh and Fisher(1990)}]{murtaugh1990bivariate}
Murtaugh, P.~A. and Fisher, L.~D. (1990).
\newblock Bivariate binary models of efficacy and toxicity in dose-ranging
  trials.
\newblock \emph{Communications in Statistics-Theory and Methods} \textbf{19},
  2003--2020.

\bibitem[{Paoletti \emph{et~al.}(2015)Paoletti, Ezzalfani, and
  Le~Tourneau}]{paoletti2015statistical}
Paoletti, X., Ezzalfani, M., and Le~Tourneau, C. (2015).
\newblock Statistical controversies in clinical research: requiem for the 3+ 3
  design for phase I trials.
\newblock \emph{Annals of Oncology} \textbf{26}, 1808--1812.

\bibitem[{Rogatko \emph{et~al.}(2008)Rogatko, Ghosh, Vidakovic, and
  Tighiouart}]{rogatko2008patient}
Rogatko, A., Ghosh, P., Vidakovic, B., and Tighiouart, M. (2008).
\newblock Patient-specific dose adjustment in the cancer clinical trial
  setting.
\newblock \emph{Pharmaceutical Medicine} \textbf{22}, 345--350.

\bibitem[{Sato \emph{et~al.}(2016)Sato, Hirakawa, and
  Hamada}]{sato2016adaptive}
Sato, H., Hirakawa, A., and Hamada, C. (2016).
\newblock An adaptive dose-finding method using a change-point model for
  molecularly targeted agents in phase I trials.
\newblock \emph{Statistics in medicine} \textbf{35}, 4093--4109.

\bibitem[{Shi and Yin(2013)}]{shi2013escalation}
Shi, Y. and Yin, G. (2013).
\newblock Escalation with overdose control for phase I drug-combination trials.
\newblock \emph{Statistics in medicine} \textbf{32}, 4400--4412.

\bibitem[{Shimamura \emph{et~al.}(2018)Shimamura, Hamada, Matsui, and
  Hirakawa}]{shimamura2018two}
Shimamura, F., Hamada, C., Matsui, S., and Hirakawa, A. (2018).
\newblock Two-stage approach based on zone and dose findings for two-agent
  combination Phase I/II trials.
\newblock \emph{Journal of biopharmaceutical statistics} , 1--13.

\bibitem[{Thall and Cook(2004)}]{thall2004dose}
Thall, P.~F. and Cook, J.~D. (2004).
\newblock Dose-Finding Based on Efficacy-Toxicity Trade-Offs.
\newblock \emph{Biometrics} \textbf{60}, 684--693.

\bibitem[{Thall and Russell(1998)}]{thall1998strategy}
Thall, P.~F. and Russell, K.~E. (1998).
\newblock A strategy for dose-finding and safety monitoring based on efficacy
  and adverse outcomes in phase I/II clinical trials.
\newblock \emph{Biometrics} , 251--264.

\bibitem[{Tighiouart(2019)}]{tighiouart2018two}
Tighiouart, M. (2019).
\newblock Two-stage design for phase I--II cancer clinical trials using
  continuous dose combinations of cytotoxic agents.
\newblock \emph{Journal of the Royal Statistical Society: Series C (Applied
  Statistics)} \textbf{68}, 235--250.

\bibitem[{Tighiouart \emph{et~al.}(2017)Tighiouart, Li, and
  Rogatko}]{tighiouart2017bayesian}
Tighiouart, M., Li, Q., and Rogatko, A. (2017).
\newblock A Bayesian adaptive design for estimating the maximum tolerated dose
  curve using drug combinations in cancer phase I clinical trials.
\newblock \emph{Statistics in medicine} \textbf{36}, 280--290.

\bibitem[{Tighiouart and Rogatko(2012)}]{tighiouart2012number}
Tighiouart, M. and Rogatko, A. (2012).
\newblock Number of patients per cohort and sample size considerations using
  dose escalation with overdose control.
\newblock \emph{Journal of Probability and Statistics} \textbf{2012}.

\bibitem[{Tighiouart \emph{et~al.}(2005)Tighiouart, Rogatko, and
  Babb}]{tighiouart2005flexible}
Tighiouart, M., Rogatko, A., and Babb, J.~S. (2005).
\newblock Flexible Bayesian methods for cancer phase I clinical trials. Dose
  escalation with overdose control.
\newblock \emph{Statistics in medicine} \textbf{24}, 2183--2196.

\bibitem[{Tighiouart \emph{et~al.}(2010)Tighiouart, Rogatko
  \emph{et~al.}}]{tighiouart2010dose}
Tighiouart, M., Rogatko, A., \emph{et~al.} (2010).
\newblock Dose finding with escalation with overdose control (EWOC) in cancer
  clinical trials.
\newblock \emph{Statistical Science} \textbf{25}, 217--226.

\bibitem[{Wages and Conaway(2014)}]{wages2014phase}
Wages, N.~A. and Conaway, M.~R. (2014).
\newblock Phase I/II adaptive design for drug combination oncology trials.
\newblock \emph{Statistics in medicine} \textbf{33}, 1990--2003.

\bibitem[{Wheeler \emph{et~al.}(2017)Wheeler, Sweeting, and
  Mander}]{wheeler2017toxicity}
Wheeler, G.~M., Sweeting, M.~J., and Mander, A.~P. (2017).
\newblock Toxicity-dependent feasibility bounds for the escalation with
  overdose control approach in phase I cancer trials.
\newblock \emph{Statistics in medicine} \textbf{36}, 2499--2513.

\bibitem[{Yin \emph{et~al.}(2006)Yin, Li, and Ji}]{yin2006bayesian}
Yin, G., Li, Y., and Ji, Y. (2006).
\newblock Bayesian Dose-Finding in Phase I/II Clinical Trials Using Toxicity
  and Efficacy Odds Ratios.
\newblock \emph{Biometrics} \textbf{62}, 777--787.

\bibitem[{Yu \emph{et~al.}(2016)Yu, Hutson, Siddiqui, and Kedron}]{yu2016group}
Yu, J., Hutson, A.~D., Siddiqui, A.~H., and Kedron, M.~A. (2016).
\newblock Group sequential control of overall toxicity incidents in clinical
  trials--non-Bayesian and Bayesian approaches.
\newblock \emph{Statistical methods in medical research} \textbf{25}, 64--80.

\bibitem[{Yuan and Yin(2009)}]{yuan2009bayesian}
Yuan, Y. and Yin, G. (2009).
\newblock Bayesian dose finding by jointly modelling toxicity and efficacy as
  time-to-event outcomes.
\newblock \emph{Journal of the Royal Statistical Society: Series C (Applied
  Statistics)} \textbf{58}, 719--736.

\bibitem[{Yuan and Yin(2011)}]{yuan2011bayesian}
Yuan, Y. and Yin, G. (2011).
\newblock Bayesian phase I/II adaptively randomized oncology trials with
  combined drugs.
\newblock \emph{The annals of applied statistics} \textbf{5}, 924.

\bibitem[{Zhang and Yuan(2016)}]{zhang2016practical}
Zhang, L. and Yuan, Y. (2016).
\newblock A practical Bayesian design to identify the maximum tolerated dose
  contour for drug combination trials.
\newblock \emph{Statistics in medicine} \textbf{35}, 4924--4936.

\end{thebibliography}

\end{document}